\documentclass{PoS}
\newcommand{\tc}{T_c}
\newcommand{\nc}{N_c}
\newcommand{\ns}{N_s}
\newcommand{\nt}{N_t}
\newcommand{\lms}{\Lambda_{\overline{\scriptscriptstyle MS}}}
\newcommand{\tclms}{T_c / \Lambda_{\overline{\scriptscriptstyle MS}}}
\newcommand{\sun}{{\rm SU}(N_c)}
\newcommand{\beq}{\begin{equation}}
\newcommand{\eeq}{\end{equation}}
\newcommand{\ept}{(\epsilon-3p)/T^4}
\newcommand{\ep}{\epsilon-3p}
\newcommand{\simgt}{\,\rlap{\lower 7.5 pt\hbox{$\mathchar \sim$}}\raise 3 pt \hb
ox{$>$}\,}
\newcommand{\simlt}{\,\rlap{\lower 7.5 pt\hbox{$\mathchar \sim$}}\raise 3 pt \hb
ox{$<$}\,}

\title{Properties of the SU($N_c$) Gluon Plasma}

\ShortTitle{SU($N_c$) Plasma}

\author{\speaker{Saumen Datta} and Sourendu Gupta \\
        Tata Institute of Fundamental Research\\
	Homi Bhabha Road, Mumbai 400005, India. \\
        E-mail: \email{saumen@theory.tifr.res.in},
                \email{sgupta@tifr.res.in}}

\abstract{We investigate the deconfinement transition in $\sun$ gauge theories,
  and properties of the deconfined phase. 
  A detailed lattice study of SU(4) and SU(6) gauge theories are conducted,
  and finite volume and cutoff effects on thermodynamic observables are
  studied. The scaling of the deconfinement transition point with
  lattice spacing is used to calculate $\lms$.  
  The continuum estimates of the thermodynamic quantities are used to study 
  properties of the gluon plasma. In particular, the approach to conformal limit
  is studied. We do not find any evidence of a strongly coupled, conformal phase
  in these theories.}

\FullConference{The XXVII International Symposium on Lattice Field Theory - LAT2009\\
		 July 26-31 2009\\
		 Peking University, Beijing, China}

\begin{document}

\section{Introduction}
At very high temperatures, strongly interacting matter is known to
exist in a deconfined, chirally symmetric state. The nature of the
transition to this state is quite sensitive to the quark sector. 
For physical quarks, one has no phase transition, only a crossover. 
For very light quarks one expects a chiral symmetry restoring transition,
whose order will depend on the mass and the number of quark flavors.
On the other hand, for pure SU(3) gauge theory, which is the limit of
QCD with infinitely heavy quarks, one has a first order, deconfining 
transition. See Ref. \cite{qm09} for further discussions of the phase diagram.

While the nature of the transition is sensitive to the quark content, one can
expect similarities in the properties of the high temperature phase of
the pure gauge theory and of QCD. In particular, while at very high
temperatures, the theories are expected to be weakly coupled due to asymptotic
freedom, at moderately high temperatures $\sim 1-3 \tc$, large
deviations from the weak coupling theory are seen in both
theories. An understanding of the interactions in the pure gauge
theory plasma will surely help in our understanding of the QCD plasma.

A study of the $\sun$ gauge theory in the limit of large $\nc$ may be
of help in this respect. Large $\nc$ arguments have been used to understand various 
features of the SU(3) deconfinement transition \cite{rob}. Also, interesting 
phase structures have been suggested for $\sun$ theory with quarks, when the baryon 
number chemical potential $\mu_B \sim \nc$ \cite{quarkonia}. Analogies have also been 
drawn between the high temperature phase of QCD and the analytically tractable, 
${\mathcal N}$ =4 supersymmetric SU($\infty$) theory \cite{gubser}.
Of course, connecting the SU($\nc \to \infty$) theory to SU(3)
will require that the corrections in $1/\nc$ are small. 
An estimate of the size of the corrections can be obtained by
studying SU($\nc$) theories with different $\nc$ on lattice.

In this paper, we report on a study of the $\sun$ gauge theories at
finite temperatures for $\nc$ = 4,6. In Sec. \ref{sec.tc} we study the
deconfinement transition. Using a finite volume analysis at different
lattice spacings, we establish the first order nature of the
transition. We also study whether asymptotic scaling holds in the transition 
regime, and use it for setting the scale through $\tclms$.
More details about the material in this section can be found in
Ref. \cite{prd}. In Sec. \ref{sec.eos} we study the equation of state of the $\sun$
gluon plasma. A comparison of the energy density and pressure for
SU(3,4,6) gives us an idea of the size of the $1/\nc$ corrections. 
In the final section we discuss the physical implications of our results. Some of these
results were earlier reported in ref. \cite{qm09}.

SU($\nc$) gauge theories have been investigated previously on the
lattice. In particular, the deconfinement transition has been studied
in Ref. \cite{gavai,teper}, while the equation of state close to $\tc$
has been studied in Ref. \cite{barak}. Our work focuses on the
continuum limit, and also explores higher temperatures in the equation
of state, which allows us to investigate the issue of approach to conformality.
Equation of state at high temperatures have also been reported in this
conference by Panero \cite{panero}, who used a coarser lattice, but
reported also results for SU(8).

\section{$\tc$ and scale setting}
\label{sec.tc}
SU(3) gauge theory is known to have a weak first order transition,
with the Polyakov loop, 
\beq
L = {1 \over V} \sum_{\vec{x}} {1 \over \nc} {\rm Tr} \ \prod_{i=1}^{N_t}
U_t(\vec{x},i)
\label{eq.pol} \eeq
acting as the order parameter. Here $V=\ns^3$ is the spatial volume and
$\nt$ is the temporal extent of the finite temperature lattice, and
$U_\mu(\vec{x},x_0)$ is the gauge link matrix connecting the sites
$x=(\vec{x},x_0)$ and $x+a \hat{\mu}$, where $a$ is the lattice spacing. 

\begin{figure}[htb]
\centerline{\includegraphics[width=.45\textwidth]{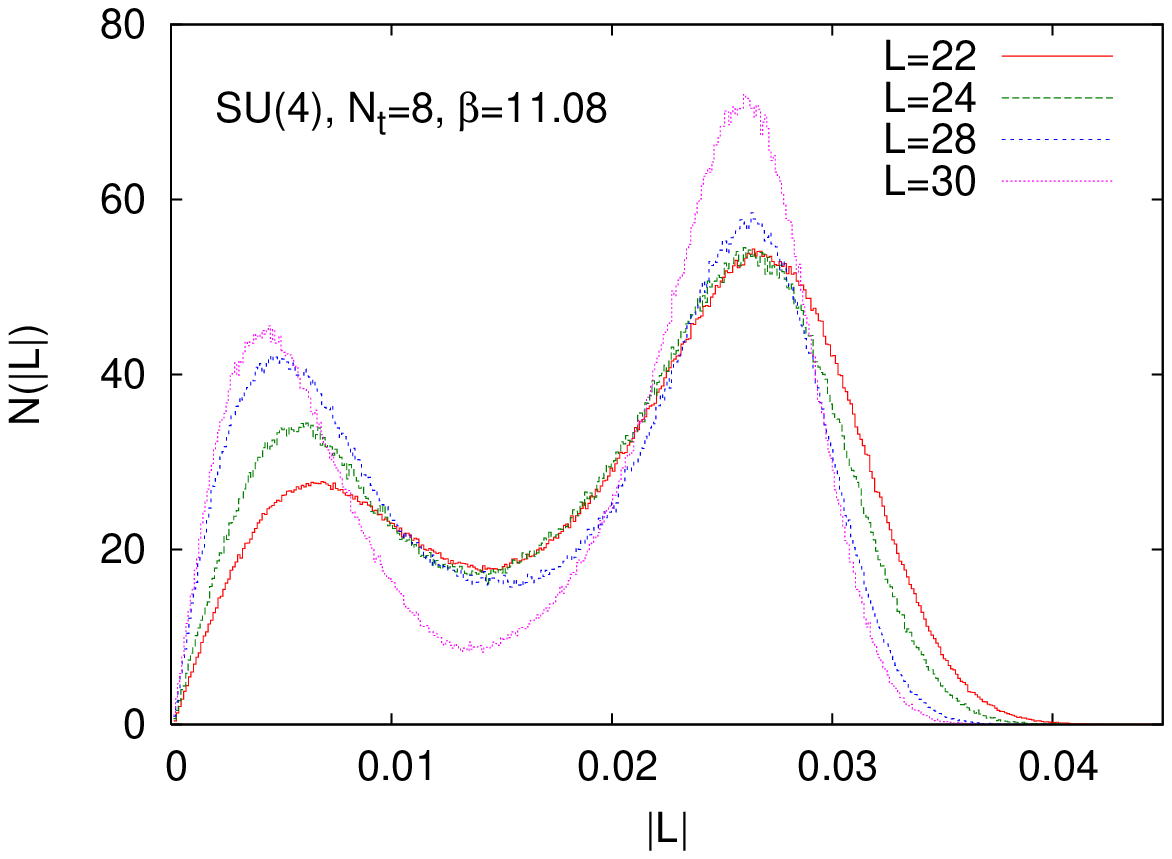}
\includegraphics[width=.45\textwidth]{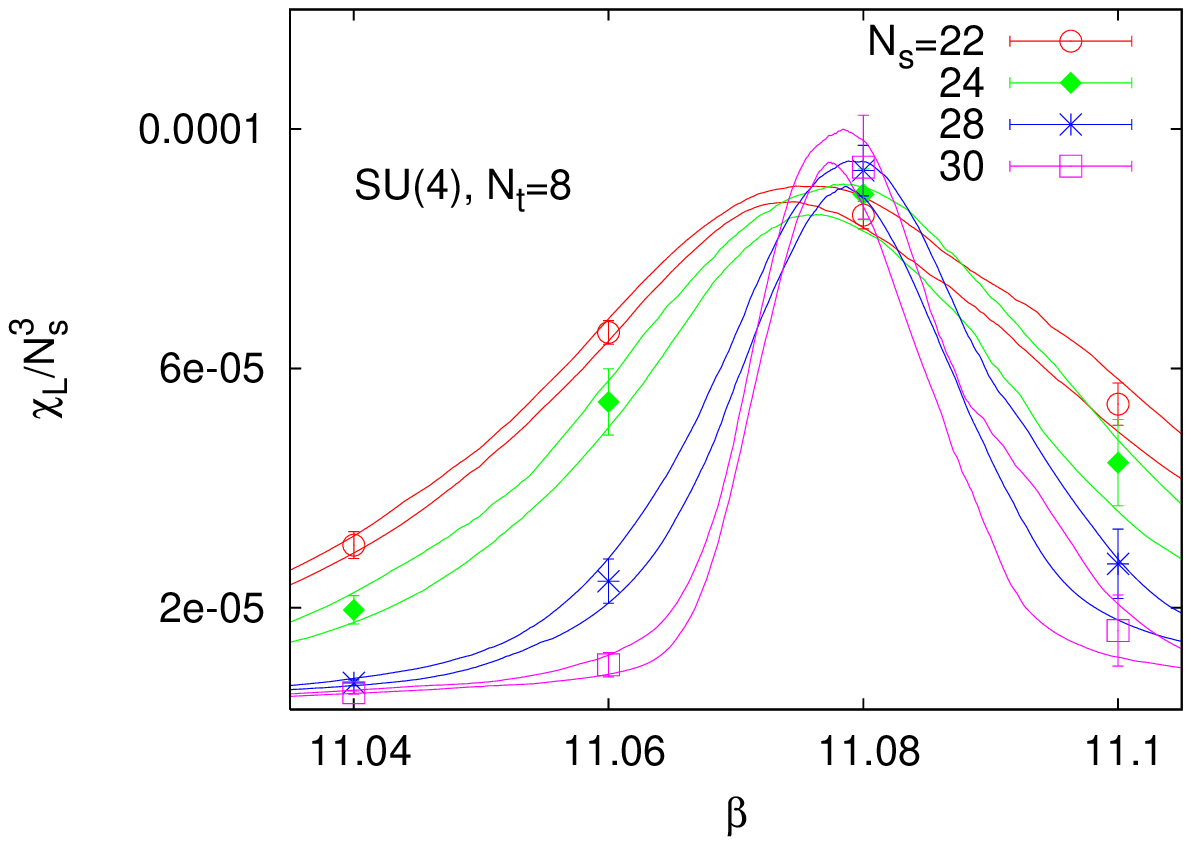}}
\caption{Finite size analysis of the deconfinement transition for
  SU(4) gauge theories, on $\nt$ = 8 lattices. (Left) Histogram of $|L|$,
  for lattices with $\ns$ = 22 - 30. (Right) Volume dependence of the $|L|$
  susceptibility.}
\label{fig.tc}
\end{figure}

We have carried out a detailed study of the deconfinement transition
in SU(4) and SU(6) gauge theories, on lattices with spacings of
$a^{-1} \leq 6 \tc$ with the Wilson action. Finite volume analysis was
carried out on lattices with $\nt$ = 6 and 8 by using various aspect ratios,
$\zeta = \ns / \nt$, in the range $\sim$ 2.5 - 4. In the transition
regime, $L$ shows a clear signal of a discontinuous transition,
tunnelling between the low temperature phase around $L$ = 0 and $\nc$
non-zero values, related by rotations of $2 \pi/\nc$, for the high
temperature phase. Figure \ref{fig.tc} shows the histogram for
$|L|$ for SU(4) gauge theory on $\nt$ = 8 lattices, showing clearly
the tunnelling between the confined and deconfined phases, with the
two-peak structure getting sharper as one goes to larger lattices. 
This is the expected behavior for a first order transition. For a more 
quantitative analysis, we look at the susceptibility of the order parameter, 
\beq
\chi_L = \ns^3 \left\{ \langle |L|^2 \rangle - \langle |L| \rangle^2
  \right\}
\label{eq.susc} \eeq
For a first order transition, $\chi_L \sim V$. In Fig. \ref{fig.tc}
we also show the result of a multihistogramming analysis of $\chi_L/V$ 
for the same set, confirming the first order nature of the
transition. More details, and figures for SU(6), can be obtained in Ref. \cite{prd}.

\begin{figure}[htb]
\centerline{\includegraphics[width=.45\textwidth]{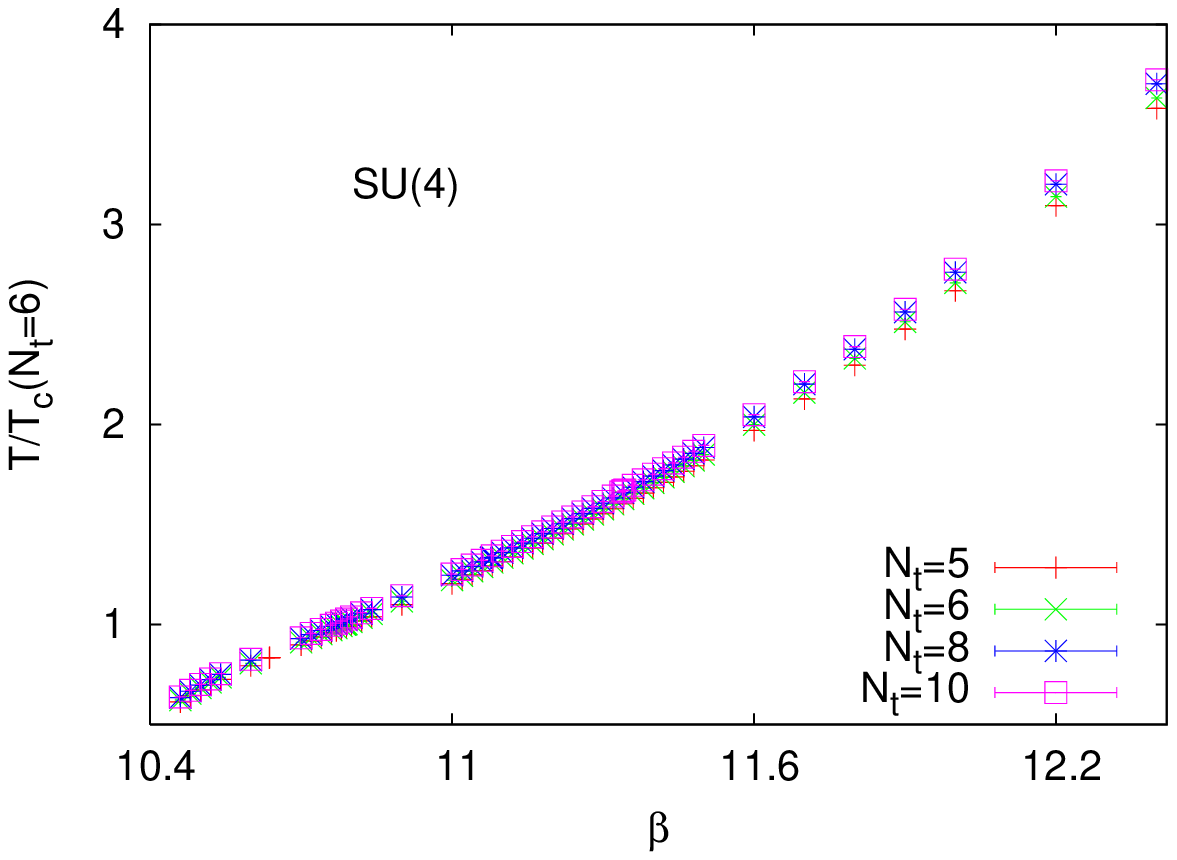}
\includegraphics[width=.45\textwidth]{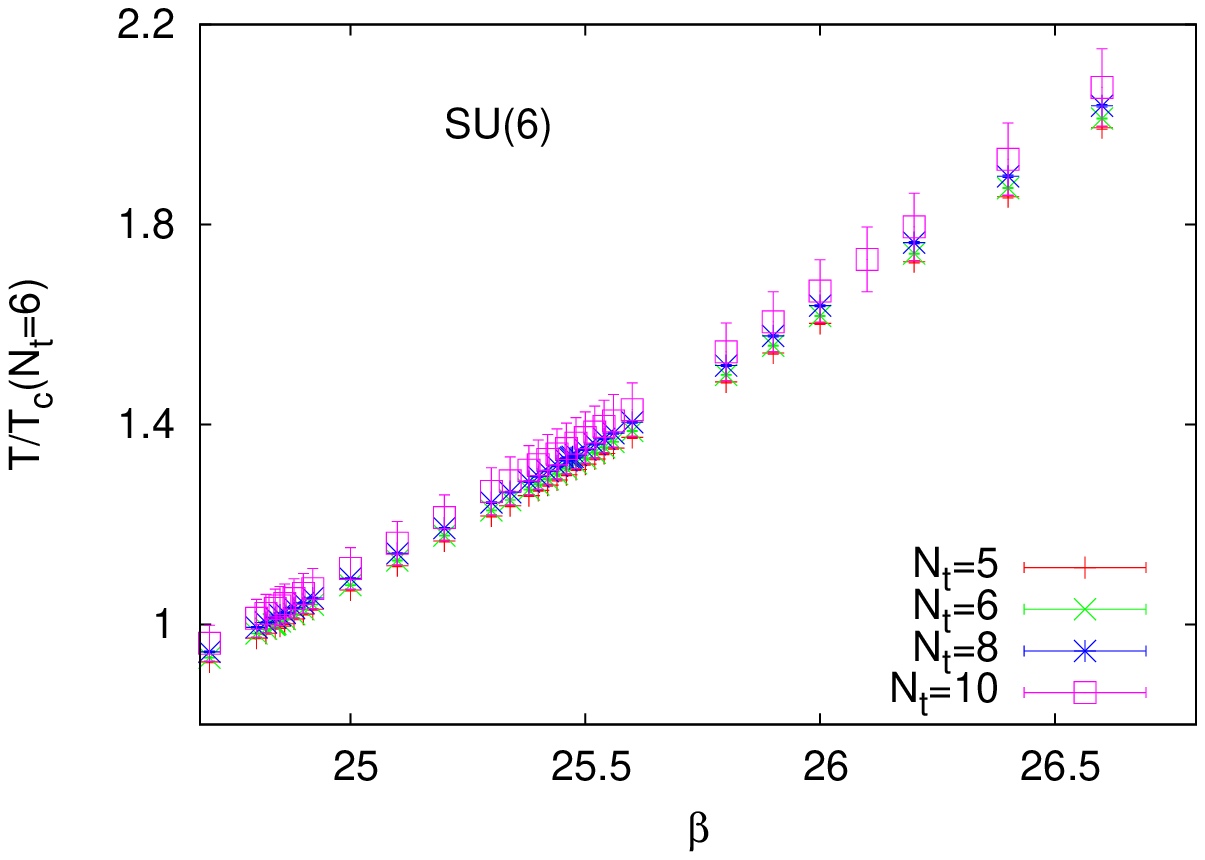}}
\caption{Temperature scale, $T/T_c (\beta)$, for $\nt$ = 6, using
  $\beta_c(\nt)$ for different $\nt$, for (a) SU(4) and (b) SU(6).} 
\label{fig.tbtc}
\end{figure}

To confirm that one is looking at a physical transition, one needs to
confirm that $\beta_c(\nt)$ has the proper scaling behavior, so that
the transition temperature, $\tc = 1/\nt a(\beta_c)$, is independent
of $\nt$. To study this scaling, we also add runs at $\nt$ = 10 for
both SU(4) and SU(6), and $\nt$ = 12 for SU(4), at a single volume
($\ns$ = 24). We use the two-loop renormalization group evolution
(RGE), and use $\beta_c(N_t)$ for the
different $\nt$ to set a temperature scale. In Fig. \ref{fig.tbtc}
we show the temperature scale for $\nt$ = 6, set by using the
$\beta_c(N_t)$ for different $\nt$, for both SU(4) and SU(6), Here 
we have also included $\beta_c(\nt=5)$, taken
from Ref. \cite{teper}. As the figure reveals, while some deviations 
from scaling are seen at smaller $\nt$, in particular for $\nt$ = 5,
very good scaling is observed for $\nt \geq$ = 8.

This allows us to use the data for $\nt \geq 8$ to get an estimate for
the scale, $\lms$, of the theory \cite{sourendu}. We use the known
expansions of plaquette to
get the renormalized coupling, and use the two-loop running to convert
it to $\lms$. Since we are performing the calculation at a finite
order of the weak coupling expansion, we also investigate the scheme
dependence, by calculating the renormalized coupling
in some other scheme, getting the $\Lambda$ parameter in this scheme, and
then converting it to $\lms$. In table \ref{tbl.lms} we show the results
of such estimates, where we have also used the E-scheme \cite{parisi}
and the V-scheme \cite{lepage}. The data for $\nc = 3$ are taken from
Ref. \cite{boyd}. We find that there is considerable scheme dependence
in the estimation of $\tclms$. We also see that within a scheme, the
dependence on the number of colors is quite weak. 

\begin{table}
\begin{center}
\begin{tabular}{|c|lll|}
\hline
$\nc$ & E-scheme & V-scheme & $\overline{\rm MS}$-scheme \\
\hline
3 & 1.19(3) & 1.12(3) & 1.20(2) \\
4 & 1.235(1) & 1.153(1) & 1.236(1) \\
6 & 1.222(1) & 1.135(1) & 1.217(1) \\
\hline
\end{tabular} \end{center}
\caption{Continuum estimate of $\tclms$ for SU($\nc$) gauge theories 
with $\nc$ = 3,4 and 6. The renormalized couplings in different
schemes are obtained from plaquette data, and converted to $a \Lambda$
using the two-loop RGE. These are converted to $\tclms$ using the
known relations between schemes. A constant fit to data for $\nt \geq
8$ is used for the continuum values.}
\label{tbl.lms} \end{table}

\section{Equation of State}
\label{sec.eos}

For estimating the thermodynamic quantities, we use the integral
method \cite{boyd}. The pressure, p, and $\ep$, where $\epsilon$ is the
energy density, are calculated from the plaquette data, \cite{boyd}
\beq
{p(T) \over T^4} - {p(T_0) \over T_0^4} = 6 \nt^4 \int_{\beta_0}^\beta d\beta
    \ \Delta P(\beta,T), \qquad \qquad \qquad
{\ep \over T^4} = 6 \nt^4 \, {\partial \beta \over \partial a} \ \Delta P(\beta,T)
\eeq
where $\Delta P(\beta,T)$ is the difference in the plaquette observables between 
the finte temperature lattice and the corresponding zero temperature 
(symmetric) lattice, calculated at the coupling $\beta$, 
$\Delta P(\beta, T) = P(\beta, T) - P(\beta,
T=0)$. $T_0$ is some reference temperature. We find that $p(T) \sim 0$
within our errors till temperatures $\sim 0.9 \tc$, and evaluate
$p/T^4$ by taking $\beta_0(T < 0.8\tc)$ as the lower limit of the integral.
For ${\partial \beta \over \partial a}$ we use the two-loop
result, with the coupling defined through the V-scheme.

\begin{figure}[htb]
\centerline{\includegraphics[width=.45\textwidth]{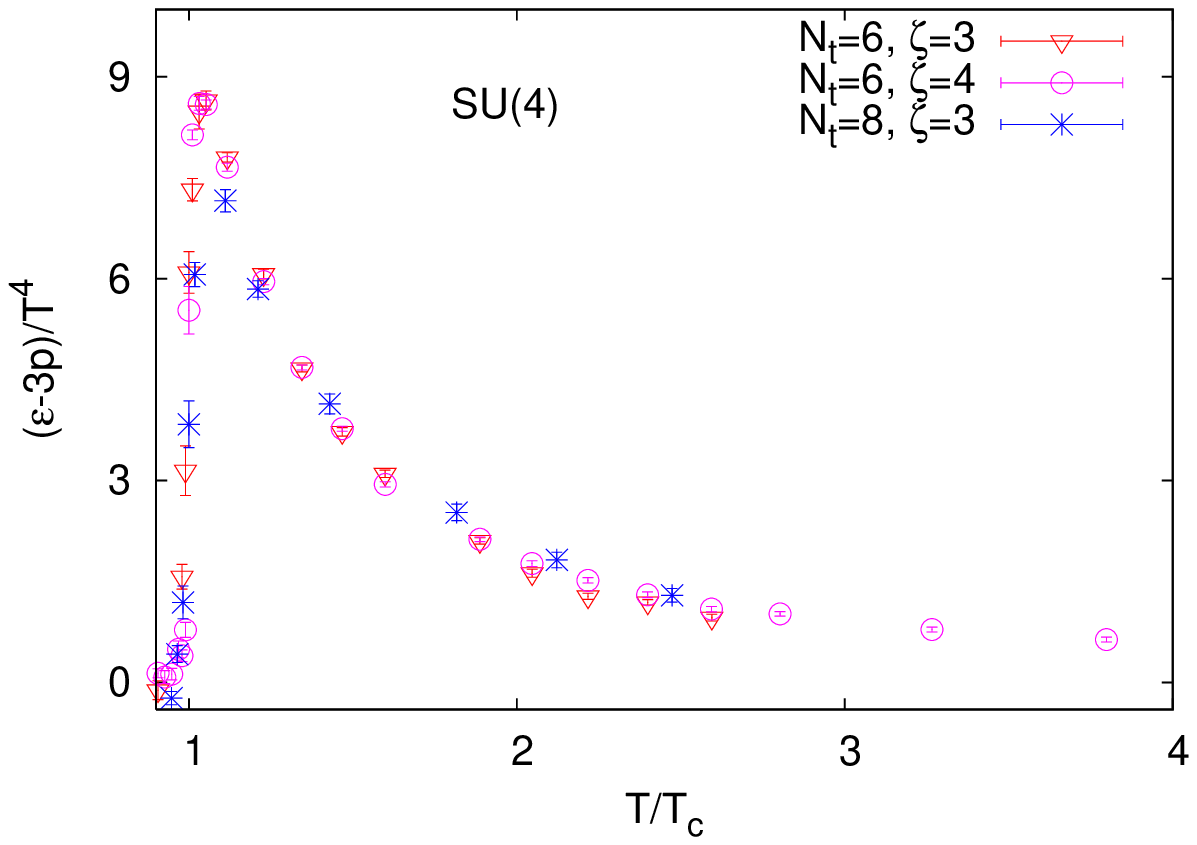}
\includegraphics[width=.45\textwidth]{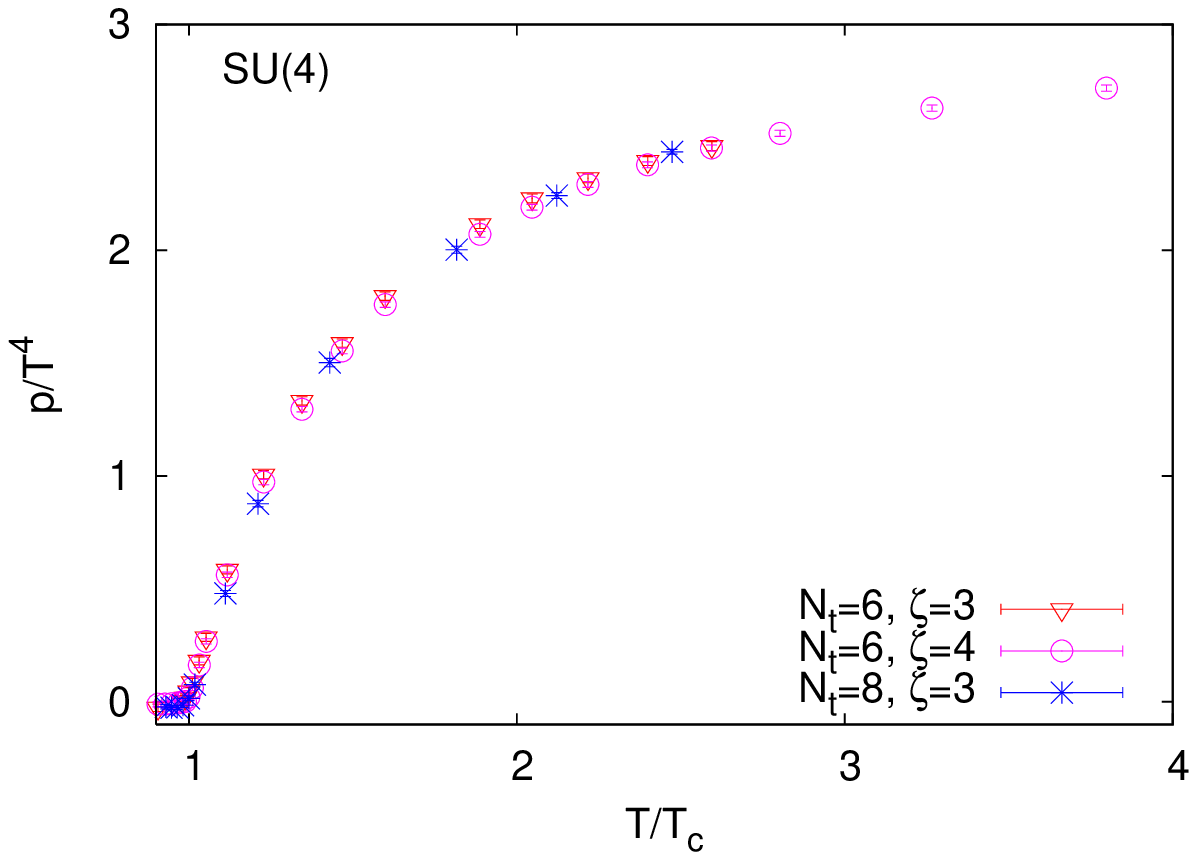}}
\centerline{\includegraphics[width=.45\textwidth]{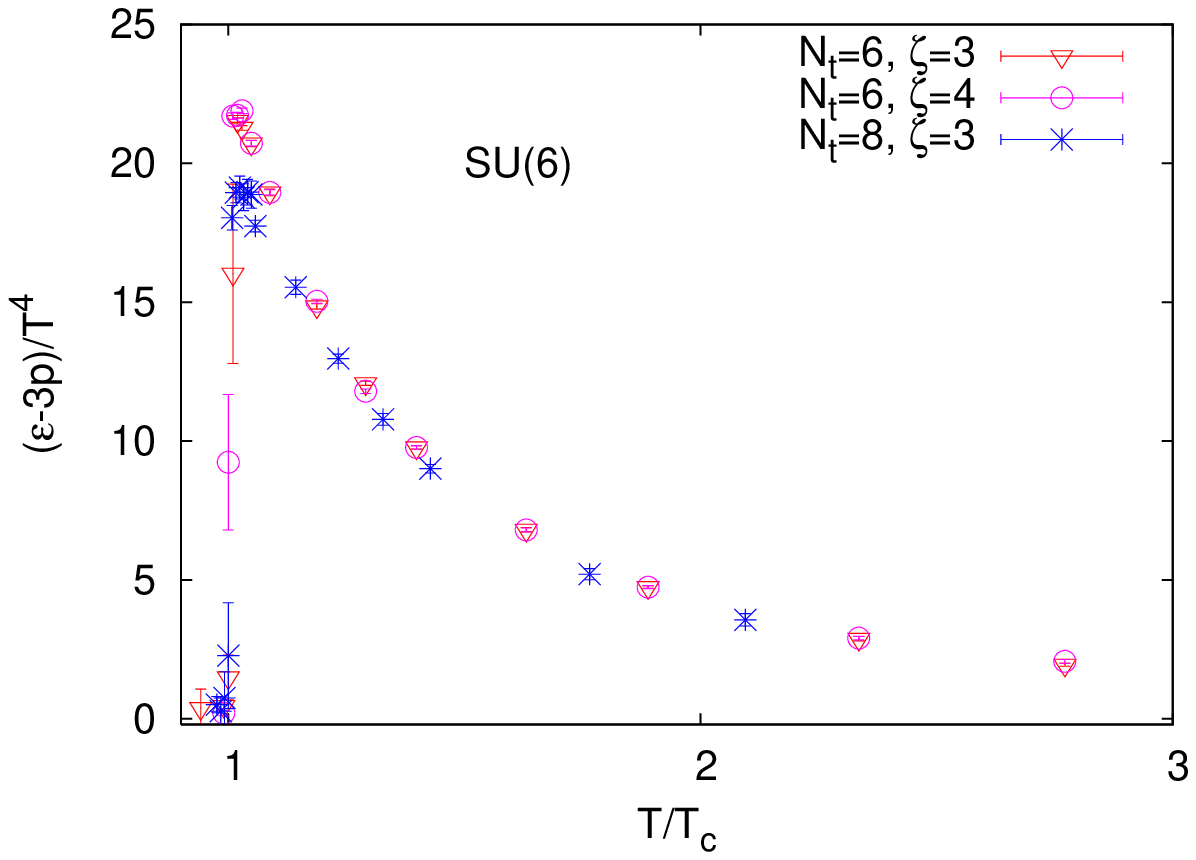}
\includegraphics[width=.45\textwidth]{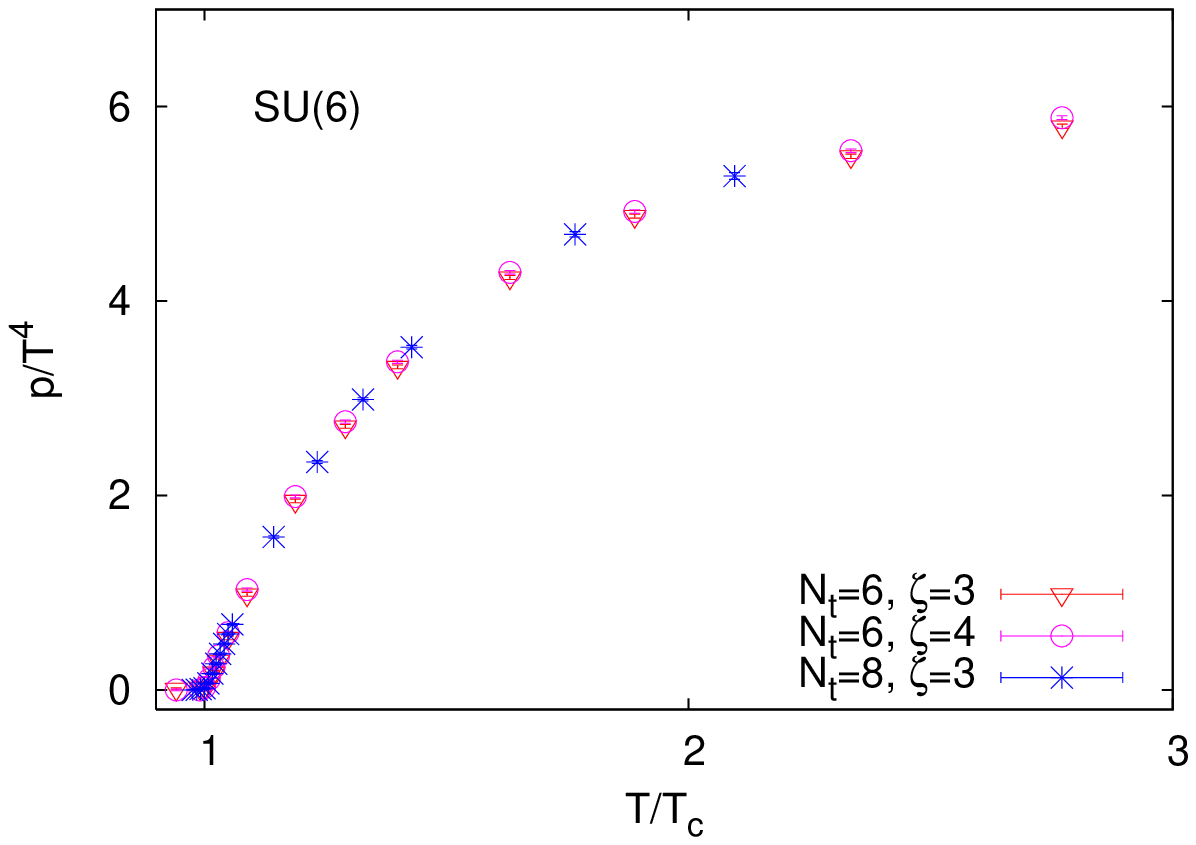}}
\caption{Volume and cutoff dependence on $\ept$ (left) and $p/T^4$ (right) for SU(4)
(top) and SU(6) (bottom) gauge theories. The results shown have been
  normalized by the known discretization error in free theory, so that
  the Stefan-Boltzmann limit for $p/T^4$ for both the $\nt$ is
  $(\nc^2-1) \pi^2/45$ (see text).}
\label{fig.sstm}
\end{figure}

The results of Sec. \ref{sec.tc} indicate that two-loop beta function
should work fine when $a \leq 1/8 \tc$, while some cutoff effect may be seen
in coarser lattices. Since we are interested in the thermodynamic and
continuum limit results, we investigate the cutoff and finite volume
effects by using two aspect ratios and two cutoffs. Note that the
Stefan-Boltzmann limit in the free theory has a known dependence on
$\nt$: $\epsilon_{SB} \ = \ 3 p_{SB} \ = \ (\nc^2 -1) \pi^2 / 15. \
R(\nt)$ where $R(\nt) = 1 + 8 \pi^2 / 21 \nt^2 + ...$ is the known
discretization error in the integral method \cite{engels,barak}.
When comparing data at different $\nt$ in Fig. \ref{fig.sstm}, we 
normalize them by this known discretization error. 

As Fig. \ref{fig.sstm} indicates, there is a considerable cutoff
effect in $\ep$ in the temperature range $\leq 1.1 \tc$ for the $\nt = 6$
lattice, for both SU(4) and SU(6). For $T \geq 1.3 \tc$ for $\nt = 6$, 
the cutoff effect is small, which is also expected since $a=1/7.8 \tc$ and is 
near the scaling regime. For pressure, the cutoff effect is already seen to be small
at $\tc$ for $\nt = 6$, for both the theories. We also see
that within the statistical uncertainties, volume dependence is small
already for an aspect ratio $\zeta = 3$.

\begin{figure}[htb]
\centerline{\includegraphics[width=.6\textwidth]{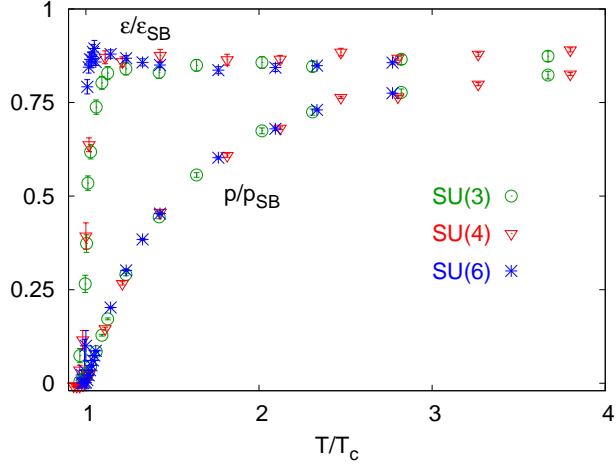}}
\caption{Results for energy density and pressure for SU(3-6) gauge
  theories, normalized to the corresponding Stefan-Boltzmann
  values. For SU(3), we have used the plaquette data of Ref. \cite{boyd}.}
\label{fig.thermo}
\end{figure}

In Fig. \ref{fig.thermo} we show the results for $\sun$
thermodynamics for $\nc$ = 3, 4 and 6. For $\nc$ = 3, we have taken
the plaquette data from Ref. \cite{boyd} and analyzed it similarly to
$\nc$ = 4 and 6. To emphasize both the deviation from
the Stefan-Boltzmann limit, and to look at evidence of corrections to
the leading $\nc^2$ behavior of the extensive quantities, we plot the
energy density and pressure normalized by their Stefan-Boltzmann
values, $\epsilon/\epsilon_{SB}$ and $p/p_{SB}$. Two comments are
in order here. First, The thermodynamic quantities differ
from their Stefan-Boltzmann values considerably even for temperatures
near 4 $\tc$. Also, the approach to the Stefan-Boltzmann limit is
slow. Second, the correction to the leading $\nc^2$ behavior is small,
as there is no systematic change between $\nc$ = 3 to 6. Note that a
subleading correction is expected to be ${\mathcal O}(1/\nc^2)$, and
therefore, four times larger in SU(3) than in SU(6). 

\section{Summary and Discussion}
\label{sec.dis}
In this report we have studied the deconfinement transition in $\sun$
gauge theories, with emphasis on the continuum results. We have
studied in detail the transition point, using a finite volume analysis
to establish the first order nature of the transition for $\nc$ = 4
and 6. We have also found that for $\nt \geq$ 8, the two-loop RGE
works quite well at the transition point. We have used this to set the 
scale of the theory, by transmuting the coupling at the deconfinement
transition, $\beta_c$, to get $\tclms$. We find this quantity to be very weakly
dependent on $\nc$. 

Next we have studied the thermodynamics of the deconfined plasma. Our
results for scaling imply that for calculating the thermodynamic
quantities near $\tc$, one needs lattices with $\nt \geq$ 8. It was
found that an aspect ratio of 3 was enough for the finite volume
effects to stay under control. 

Our estimates for the continuum thermodynamic quantities are shown in
Fig. \ref{fig.thermo}. The figure denotes that the correction to the
leading $\nc^2$ contribution is rather small in both energy density
and pressure, already for $\nc$ = 3. It is also found that the energy
density and pressure differ significantly from their Stefan-Boltzmann
values, even deep in the deconfined phase. 

\begin{figure}[htb]
\centerline{\includegraphics[width=.6\textwidth]{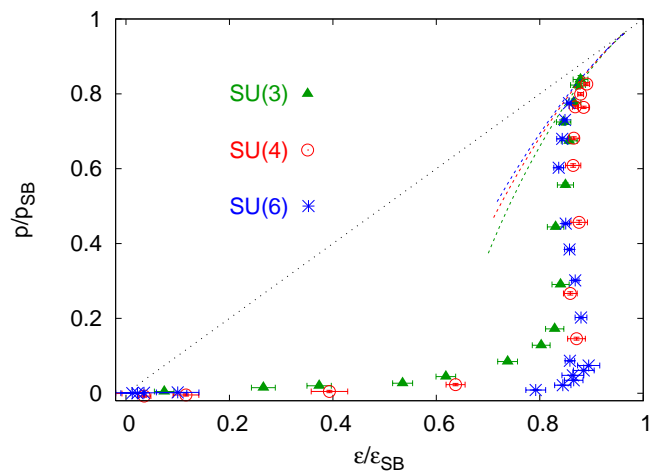}}
\caption{Approach to conformality in $\sun$ gauge theories. Also shown
  are the weak coupling results from \cite{mikko}.}
\label{fig.conform}
\end{figure}

The large deviation from the Stefan-Boltzmann value, and the fact that
some strongly coupled conformal field theories show similar deviations
from Stefan-Boltzmann limit \cite{gubser}, have sometimes been used in the
literature to speculate about a strongly coupled, conformal regime in 
pure $\sun$ gauge theories. The thermodynamics results can be used to
investigate the feasibility of such a phase. Following
Ref. \cite{swagato}, in Fig. \ref{fig.conform} we plot the energy
density vs. pressure, normalized by the corresponding Stefan-Boltzmann 
values. By construction the point at (1,1) is the Stefan-Boltzmann
limit, while the diagonal line denotes conformality. Also shown are
the weak coupling lines for the theories with the different number of
colors \cite{mikko}. We find that the weak coupling line is reached
before the conformal line, indicating the absence of a strongly
coupled, conformal phase in the $\sun$ gluon plasma.

We thank Mikko Laine for providing us with the weak
coupling results of Fig. \ref{fig.conform}. The computations were
carried out on the workstation farm of the department of theoretical
physics, TIFR. We thank Ajay Salve for technical support.

\end{document}